\documentclass[prl,floatfix,twocolumn]{revtex4}
\usepackage{amssymb,graphics,graphicx,times,bm,bbm,multirow}

\usepackage{times,amsmath,amsfonts,amssymb,latexsym}
\usepackage{graphicx,epsf,epsfig,verbatim}

\usepackage{epstopdf}
\usepackage{setspace}

\begin{document}

\title{Continuous Measurement of a Non-Markovian Open Quantum System}

\author{A. Shabani, J. Roden, and K. B. Whaley}

\affiliation{Berkeley Center for Quantum Information and Computation, Berkeley, California 94720 USA\\ Department of Chemistry, University of California, Berkeley, California 94720 USA}

\begin{abstract}
Continuous quantum measurement is the backbone of various methods in quantum control, quantum metrology, and quantum information. Here, we present a generalized formulation of dispersive measurement of a complex quantum systems. We describe the complex system as an open quantum system that is strongly coupled to a non-Markovian environment, enabling the treatment of a broad variety of natural or engineered complex systems. The system is monitored via a probe resonator coupled to a broadband (Markovian) reservoir. Based on this model, we derive a formalism of \textit{Stochastic Hierarchy Equations of Motion} (SHEM) describing the decoherence dynamics of the system conditioned on the measurement record. Furthermore, we demonstrate a spectroscopy method based on weak quantum measurement to reveal the non-Markovian nature of the environment, which we term \textit{weak
spectroscopy}.
\end{abstract}

\maketitle

\textit{Introduction --} Generalized or weak quantum measurement has become increasingly important in the last decade due to its application in quantum feedback control \cite{Wiseman:book,Shabani:08}, quantum metrology \cite{Wiseman:book}, quantum information \cite{Nielsen:book,Lidar:book,Blais:04}, and 
the study of quantum-classical transitions \cite{Habib:06,Everitt:05}. The existing theories consider continuous weak measurement of simple open quantum systems with Born-Markov decoherence models \cite{Blais:04,Gambetta:08,Boissonneault:09,DohertyMabuchi}. However, there is a lack of theoretical formalism to extend the exceptional capacities of weak measurement method for system identification and control to complex natural \cite{Mohseni:book} or engineered \cite{DWave,Ion} systems, i.e., systems that are large, possibly disordered and interacting strongly with their environment. The present letter addresses the demand for such advanced theories.

Cavity quantum electrodynamics (CQED) is a well established paradigm to implement weak quantum measurement protocols \cite{DohertyMabuchi, Blais:04}.
In this paper, we develop a CQED theory for continuous measurement of an arbitrary quantum system coupled to a bosonic environment. In this framework, we derive a set of coupled stochastic differential equations, SHEM, that describes the system conditional evolution in the presence of non-Markovian and possibly strong decoherence effects. As an application of our theory, we propose a simple experimental  spectroscopic procedure to diagnose the non-Markovian nature of the decoherence dynamics via continuous measurement.
Our theory can be applied
in any frequency regime, given the appropriate parameters setting. 
We shall therefore use the term CQED to refer to both cavity (optical) and circuit (microwave) QED systems.

We should emphasize that the 
current SHEM formalism describes a different measurement paradigm than the measurement interpretations of non-Markovian stochastic Schr\"odinger equations \cite{SSE1,SSE2,SSE3,SSE4,SSE7,SSE8}. The latter, originally developed to model decoherence dynamics in presence of a bosonic environment \cite{SSE1,SSE2,SSE3,SSE4}, under certain conditions, 
describe direct monitoring of the environment \cite{SSE7,SSE8}. In the measurement setting considered in this letter, we avoid such a direct environmental monitoring, rather we use an independent cavity coupled to a broadband reservoir to probe only the system.

In CQED, a quantum harmonic oscillator acting as the measurement probe interacts with the system.  
In the optical regime the oscillator is realized by a single mode of a cavity with mirror walls \cite{DohertyMabuchi}, in the microwave regime by a 1D \cite{Blais:04} or 3D \cite{Paik:11} cold electrical resonator. The oscillator, which we shall refer to from now on as the cavity, is designed to be imperfect and to have photon loss. The leaked photons carry information about the phase and amplitude of the cavity mode,
quantities that per se encode some system information. An appropriate detection scheme can therefore indirectly extract some system information by measuring the leaked photons. The photon detection occurs in real-time, leading to a continuous measurement of the quantum system. In the following, we present a full formulation of the measurement scenario for a quantum system with an arbitrary internal structure, that is coupled to a cavity mode and is
also free to interact with an additional bosonic environment. Physical examples of such systems include double quantum dots probed by a microwave resonator \cite{Frey:12}, superconducting qubits with undesired coupling to 3D cavity modes \cite{Paik:11,Ma:12}, and atoms in an optical resonator \cite{DohertyMabuchi}. Note that gaussian fluctuations with fermionic nature can be also effectively modeled as a bosonic bath \cite{Frank,Weiss:Book}. 

\begin{figure}[tp]
\includegraphics[width=8cm,height=2.8cm]{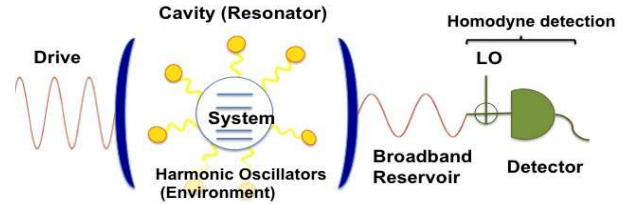}
\caption{A single mode cavity resonator probes an quantum system coupled to a bosonic environment. A detector continuously measures the photons leaking out of the cavity that carry information about the system dynamics.} \label{cavityplot}
\end{figure}

\textit{A complex system coupled to a cavity --} Consider a quantum system with Hamiltonian $H_S$ coupled to a single cavity mode with Hamiltonian $H_C=\omega_c a^\dagger a$ at frequency $\omega_c$, where $a$ is the mode lowering operator. We consider a system-cavity coupling $H_{int}=\hat{\mu}(a+a^\dagger)$ for some Hermitian system operator $\hat{\mu}$. This model captures a range of couplings, including electric dipole-electric field, spin-magnetic field, or more complicated engineered couplings \cite{Blais:04,Doherty:99}. The total system-cavity Hamiltonian is then
\begin{equation}
H_{SC}=H_S+\omega_ca^\dagger a+\hat{\mu} (a+a^\dagger)
\end{equation}
A desired system-probe coupling suitable for weak measurement has the form $O_S a^\dagger a$ \cite{Wiseman:book,Breuer:book}. One can then measure the observable $O_S$ by direct measurement of the cavity phase conjugate to the operator $a^\dagger a$. 
In the dispersive regime where the cavity is relatively far off detuned from the system resonance frequencies, i.e., for $|\hat{\mu}_{jk}=\langle j| \hat{\mu}|k\rangle |\ll |\omega_c-(\Omega_k-\Omega_j)|$ with the spectral decomposition $H_S=\sum_j \Omega_j |j \rangle\langle j|$,
the Hamiltonian $H_{SC}$ can be turned into this desired form by applying a generalized dispersive transformation $U_{D}=\exp[Xa^\dagger -X^\dagger a]$, where $X=\sum_{jk} \frac{\hat{\mu}_{jk}}{\omega_c+\Omega_j-\Omega_k} |j\rangle\langle k|$ and all terms up to second order in $X$ are retained.  Applying the rotating wave approximation and neglecting two photon creation and annihilation processes results in the dispersive Hamiltonian
\begin{equation}
H^D_{SC}=U_DH_{SC}U_D^\dagger\approx H_S^D+\omega_ca^\dagger a+O_S a^\dagger a\label{Hsc}
\end{equation}
with modified system Hamiltonian $ H_S^D=H_S-\frac{1}{2}(X^\dagger \hat{\mu}+\hat{\mu} X)$.
Now the effective system-cavity coupling $O_S a^\dagger a$ is of the desired form, with $O_S=\frac{1}{2}[\hat{\mu},X^\dagger-X]$.
The system operator $O_S$ serves as an adjustment to the cavity frequency, i.e. $\omega_c^D=\omega_c+O_S$, and therefore measuring the phase of the leaking photons reveals information about the system. Such measurement is not generally non-demolition unless $[H^D_S,O_S]=0$ \cite{Breuer:book}. We will see later that $O_S$ is the primary but not the only component of the total observable. 

We can now induce dynamics by a classical cavity drive with multiple frequency components, i.e. $\mathcal{E}(t)=\mathcal{E}_pe^{-i\omega_p t}+\sum_q \mathcal{E}_qe^{-i\omega_q t}$. The 
detuning between system and cavity in the dispersive regime allows these to be individually addressable. The component $\mathcal{E}_p$ with frequency $\omega_p$ that is near resonance with $\omega_c$ drives the cavity, while choosing $\{\omega_q\}$ close to the system frequencies allows the components $\{\mathcal{E}_q\}$ to excite the system.
The Hamiltonian for such a multi-frequency drive is
\begin{eqnarray}
H_{drive}^D=\mathcal{E}(t) U_D a^\dagger U_D^\dagger+h.c.\approx\hspace{1.2 in}\notag\\
\mathcal{E}_p e^{-i\omega_p t} a^\dagger(1+\Lambda)- \sum_q\mathcal{E}_qe^{-i\omega_{q} t}X^\dagger+h.c.\label{drive}
\end{eqnarray}
and is added to $H^D_{SC}$. Here $\Lambda=\frac{1}{2}[X^\dagger,X]$, and {\it{h.c.}} stands for Hermitian conjugate.

\textit{System decoherence and cavity leakage --} The dynamics of the system and cavity is 
further influenced by two sources of ambient interactions: a broadband electromagnetic reservoir $R$ that couples to the cavity, causing photon leakage, and an environment $E$ that induces decoherence via its coupling to the system, see fig.(\ref{cavityplot}). 
The latter could be, e.g., acoustic phonons in a quantum dot \cite{qdots} or intramolecular vibrational modes of chromophores \cite{Ishizaki:10}.  A master equation description of these process can be derived in the Born (weak coupling)-Markov (no-memory) regime \cite{Wiseman:book,Gambetta:08,Boissonneault:09}. Here we present a detailed microscopic derivation of these processes with neither Markovian nor perturbative approximation.

We shall index the modes of the electromagnetic reservoir by $0r$ and the modes of the environment by $mr$ with $m \geq 1$.  The environment $E$ is then treated as a sum of local baths of harmonic oscillators, fig.(\ref{cavityplot}), with Hamiltonian $H_{E}=\sum_{m\ge 1,r} \omega_{mr}  b_{mr}^\dagger  b_{mr}$, where $b_{mr}$ denotes the lowering operator of the $(mr)$th mode at frequency $\omega$. 
The system-environment coupling is of the 
form $H_{SE}=\sum_{m\ge 1,r}g_{mr}(b_{mr} +b_{mr}^\dagger)S_m$.
for some system operators $S_m$ and coupling strengths $g_{mr}$. 
The effect of such a bath is usually captured by its spectral density $J_m(\omega)=\sum_r g_{mr}^2\delta(\omega-\omega_{mr})$.
We model the photon leakage process by including a reservoir of electromagnetic modes, $H_{R }=\sum_{r} \omega_{0r}  b_{0r}^\dagger  b_{0r}$, that is also linearly coupled to the cavity mode: $H_{CR}=\sum_{r}g_{0r}(b_{0r}+b_{0r}^\dagger)S_m$. 
We emphasize that coupling to the electromagnetic reservoir is desirable, since this allows some photons and therefore some information to leak out of the cavity, while coupling to the environment is usually undesirable since this destroys the coherence needed for quantum information processing \cite{Nielsen:book}.

Now, we have to solve for the dynamics of a 4-component system, $S,C,E,$ and $R$ which possesses only 2-body interactions  in the laboratory frame ($H_{SC},H_{SE},H_{CR}$), but the transformation into the dispersive frame generates some perturbative 3-body  couplings. Specifically, the Hamiltonian $H_{CR}$ becomes
\begin{eqnarray}
H_{CR}^D=\sum_rg_{0r}(b_{0r}+b_{0r}^\dagger )F_0, \label{Res}
\end{eqnarray}
with $F_0=(a+a^\dagger)(1+\Lambda)-(X+X^\dagger)$.

The term $(b_{0r} +b_{0r}^\dagger )(X+X^\dagger)$ in Eq.(\ref{Res}) shows that the dispersive transformation has introduced a new reservoir channel for system decoherence, together with higher order system-cavity-reservoir terms $b_{0r}a^\dagger\Lambda$, Such modification of the system decoherence channels leads to a Purcell type effect in the dispersive regime \cite{DohertyMabuchi}. The change of frame also modifies the interaction between the system and the environment, yielding
\begin{eqnarray}
H^D_{SE}=\sum_{m\ge 1,r} g_{mr}(b_{mr}+b_{mr}^\dagger )(\overbrace{\tilde{S}_m+Q_ma^\dagger a}^{=F_m}+G_m)\label{HDSC}
\end{eqnarray}
with $\tilde{S}_m=S_m-\frac{1}{2}\{X^\dagger X,S_m\}+X^\dagger S_mX$, $Q_m=(\mathcal{D}[X]+\mathcal{D}[X^\dagger])S_m$, and $G_m=-[X^\dagger,S_m]a+[X,S_m]a^\dagger$. The superoperator $\mathcal{D}$ is defined as $\mathcal{D}[A]B=ABA^\dagger-\frac{1}{2}\{A^\dagger A,B\}$. 
The operator $\tilde{S}_m$ represents the effective system and environment coupling. We interpret the second term with the $Q_m$ operator as a cavity-state dependent decoherence of the system that will show up explicitly in the master equation.
We analyze the dynamics in the frame rotating with $H_C$. Accordingly, the last term in Eq. (\ref{HDSC}),
which contains a sum over contributions $g_{mr}(b_{mr}+b_{mr}^\dagger )G_m$ for each environmental mode 
$mr$, becomes 
\begin{eqnarray}
\sum_{m\ge 1,r}g_{mr}(b_m+b_m^\dagger)(-[X^\dagger,S_m]e^{-i\omega_c t}a+[X,S_m]e^{i\omega_c t}a^\dagger)\label{neg}
\end{eqnarray}

Here we assume the environment has negligible modes at the cavity frequency 
$\omega_c$, i.e., $J_m(\pm \omega_c)\approx 0$. This is a legitimate assumption when the environmental cutoff frequencies are smaller or comparable to the cavity-system detuning, which can be the case for e.g.,  phonons in quantum dots \cite{qdots} and vibrational modes of proteins in photosynthetic complexes \cite{Ishizaki:10}. Neglecting Eq. (\ref{neg}), yields the following Hamiltonian for the system-environment and cavity-reservoir
interactions 
\begin{eqnarray}
H^D_{SE}+H^D_{CR}=\sum_{m\ge 0,r}g_{mr}(b_{mr}+b_{mr}^\dagger ) F_m. \label{H_B}
\end{eqnarray}
The full $(S,C,R,E)$ Hamiltonian is then $H^D_{SC}+H^D_{drive}+H^D_{SE}+H^D_{CR}$. Incorporation of Eq. (\ref{neg}) will only perturbatively modiy the decoherence dynamics, while its non-Markovian effect on the cavity mode is dominated by the strong Markovian leakage process 
that is described by $H^D_{CR}$.
Tracing out the surrounding bosonic fields, $R$ and $E$, yields a combined description of the photon leakage and decoherence dynamics. The common approach to describe such open system processes is to apply the Born-Markov approximation \cite{Gambetta:08,Boissonneault:09}.
Here we take a different approach that avoids these approximations and relies instead only on assuming Gaussian statistics for the fluctuations of the bosonic fields. 
This is the approach of the hierarchy equations of motion (HEOM) \cite{Tanimura:89}, which we employ here with the Drude-Lorentz form environmental spectral density, $J_m(\omega)=2\lambda_m\gamma_m\frac{\omega}{\omega^2+\gamma_m^2}$ with coupling strength $\lambda_m$ and cut-off frequency $\gamma_m$.

The HEOM approach has recently been generalized to include arbitrary parameterized bath correlation functions for both bosonic and fermionic environments \cite{Jin:07,Ma:12}. As an exact solution, HEOM can treat any level of non-Markovianity or system-bath coupling strength and is therefore is a powerful tool for dynamical simulation of quantum systems with complex environmental interactions.
HEOM has produced successful results in modeling various systems such as quantum dots \cite{Zheng:08}, nano-devices \cite{Jin:10}, Kondo systems \cite{Li:12}, and photosynthetic complexes \cite{Ishizaki:pnas}.
The full HEOMs 
describing the system dynamics with all leakage and decoherence processes included are described in the appendix. We emphasize that our analysis employs an exact treatment of the influence of the
(non-Markovian) environment $E$ while treating the reservoir $R$ as a broadband (Markovian) field.  This description is well suited to the properties of many current implementations of optical or microwave resonators. While non-Markovian cavity leakage processes have been proposed \cite{Zhang:13}, from the perspective of quantum measurement there is however no known benefit from such generalization. 

\textit{Continuous Measurement --} Eq.(\ref{Hsc}) shows that coupling to the system modifies the effective frequency of the cavity. Consequently, photons leaving the cavity carry information about the system that is encoded in their phase. 
Measurement of the phase of the photons and hence indirect measurement of the system may be made by homodyne detection \cite{Wiseman:book,Gambetta:08}. A continuous flow of photons thereby allows us to continuously monitor the system.
The noisy current of a detector $I(t)=d\mathcal{Q}/dt$ with efficiency $\eta$ can be described by \cite{Wiseman:book,Gambetta:08}
\begin{eqnarray}
d\mathcal{Q}\propto 2\eta\kappa\langle e^{-i\phi}a+e^{i\phi}a^\dagger\rangle dt+\sqrt{2\eta\kappa}dW_t \label{current}
\end{eqnarray}
($\langle .\rangle$ means expectation value)
for a local oscillator with phase $\phi$. The parameter $\kappa$ is the cavity leakage rate determined by $J_0(\omega)$, and the infinitesimal increment $dW_t$ represents a Wiener process. 
The latter results in a measurement back-action term in the system-cavity dynamics  (see below) that is stochastically conditioned on the instantaneous outcome at the detector. In the case of Markovian leakage, such random changes allow a stochastic unravelling of the corresponding Markovian master equation \cite{Wiseman:book,Gambetta:08}. We show in appendix B that such an unravelling is possible even in the presence of a non-Markovian decoherence process described by the HEOM.

In order to ensure that the detection information reflects only the quantum state of the system, we shall consider the bad cavity regime in which the cavity state slavishly follows the system \cite{Doherty:99,Hutchison:09}. In this regime the leakage time $\kappa^{-1}$ is smaller than the time scale of the system dynamics and we may adiabatically eliminate the cavity mode, in both optical \cite{Doherty:99} and microwave \cite{Hutchison:09} regimes 
(details are given in appendix C). A good criterion for applicability of the bad cavity parameter regime is $\kappa\gg ||O_S||_1(1+|\alpha|^2)$, where $|\alpha=\mathcal{E}_p/i\kappa\rangle$ is the bare cavity coherent state for $\omega_c=\omega_p$.
This adiabatic elimination leads to our main result, which is a set of stochastic hierarchy equations of motion (SHEM) that describe the dynamics of a quantum system under continuous measurement with non-perturbative coupling to a non-Markovian bath. 
For temperatures $\beta^{-1}$ larger than $\gamma_m$, a particularly simple form is found, namely 
\begin{eqnarray}
d\sigma^{{\bf{n}}}=\mathcal{L}_{S}^{\bf{n}}[\sigma^{{\bf{n}}}]dt-\sum_{m\ge 1}n_{m}\gamma_{m}\sigma^{{\bf{n}}}dt-i\sum_{m\ge 1}[\tilde{F}_m,\sigma_{n_{m}+1}]dt\hspace{.2 in}\notag\\
\hspace{.3 in}+\sum_{m\ge 1}n_{m}(\frac{2i\lambda_m}{\beta}[\tilde{F}_m,\sigma_{n_{m}-1}]+\lambda_m\gamma_m \{\tilde{F}_m,\sigma_{n_{m}-1}\})dt\hspace{.2 in}\notag\\
-\sqrt{\frac{2\eta}{\kappa}}\mathcal{H}[i\alpha e^{-i\phi}O_S]\sigma^{{\bf{n}}} dW_t \label{SHEM}\hspace{1.7 in}
\end{eqnarray}
corresponding to the detector record
\begin{eqnarray}
d\mathcal{Q}\propto 2\eta|\alpha|\langle \bar{O}_S\rangle dt+\sqrt{2\eta\kappa}dW_t\hspace{.75 in}\label{observe}\\
\bar{O}_S=\sin(\phi-\arg(\alpha))O_S +\kappa\cos(\phi-\arg(\alpha))\Lambda\label{obs}
\end{eqnarray}
with 
$\mathcal{H}[A].=A. -Tr[A.].+h.c.$, $\tilde{F}_m=\tilde{S}_m+Q_m|\alpha|^2$, and $\mathcal{L}_{S}^{\bf{n}} .=-i[H_S^D- \sum_q(\mathcal{E}_qe^{-i\omega_{q} t}X^\dagger+h.c.)+|\alpha|^2O_S,.]+(|\alpha|^2/\sum n_{m}\gamma_{m})\mathcal{D}[O_S].+\kappa \mathcal{D}[X].$  Here $A.$ denotes action of the map A on an operator. 
The general form of the SHEM for arbitrary temperature and $\omega_c\neq\omega_p$ is derived in the appendix.

The subscript $\bf{n}$ is a matrix of non-negative indices $n_{ma}$ that defines one tier 
of the full coupled set of equations. The system density matrix
corresponds to index zero, $\rho_{S}=\sigma_{\bf{n=0}}$ and the rest of operators $\sigma_{\bf{n}\neq 0}$ are auxiliary operators that capture non-Markovian effects. The index $n_{ma}\pm 1$ denotes increase or decrease of the index $n_{ma}$. In practice, one truncates Eq.~(\ref{SHEM}) at finite indices $\{n_{ma}^t\}$ when $\sum_{m\ge 1,a}n_{ma}^t\gamma_{ma}$ is much larger than the system frequencies. Therefore, the SHEM can capture a higher level of non-Markovianity merely by including more auxiliary operators $\sigma_{\bf{n}}$. Notice that the strength of $SE$ couplings $\lambda_m$ is an unconstrained parameter that enables SHEM to handle strong decoherence. Another interesting feature is the non-Markovian nature of the measurement back-action that derives from $Q_m|\alpha|^2$.

The measured observable $\bar{O}_S$ is not limited to the result of the direct system-cavity coupling, i.e., to $O_S$. 
An additional component $\Lambda$ appears as a result of the 3-body system-cavity-reservoir coupling becoming effective in the dispersive 
frame (see above). In Ref.\cite{Shabani:prep}, we discuss how the effective observable $\bar{O}_S$ can be engineered by varying the cavity frequency $\omega_c$, the local oscillator $\alpha$, or the cavity drive phase $\phi$. That enables continuous quantum state tomography \cite{Shabani:prep}.

\textit{Weak spectroscopy for detection of non-Markovian dynamics --}
Spectroscopy can provide useful dynamical information about atomic and molecular systems by analyzing correlations in measurement outcomes. We show here that the non-destructive nature of weak measurement can empower spectroscopy, in particular, that it allows detection of the non-Markovian nature of the decoherence dynamics.  We term this general approach ``Weak Spectroscopy" \cite{WS}.

Consider a single qubit driven in resonance with amplitude $\Omega_R$ while the observable $O_S=\chi \sigma_z=\chi (|0\rangle\langle 0|-|1\rangle\langle 1|)$ is continuously measured. With Markovian decoherence, the power spectral density of the detector current \cite{PSD},
$S(\omega)$, is bell shaped with a peak around $\Omega_R$. 
\begin{figure}[tp]
\includegraphics[width=9cm,height=6cm]{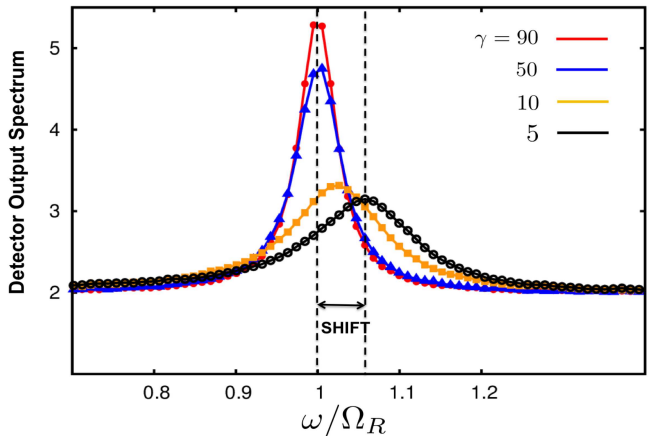}
\caption{
A qubit coupled to a non-Markovian environment 
undergoes Rabi oscillations on driving at frequency $\Omega_R$. The plot shows the 
spectrum of the detector current that continuously measures the qubit population. 
The spectrum is monotonically shifted and broadened as the environment bandwidth $\gamma$ decreases. 
Due to the finite time range over which the detector current is calculated, the spectrum has a finite frequency resolution, i.e., it consists of discrete points.} \label{Pspectrum}
\end{figure}
A non-Markovian environment modifies the detector spectrum. In the SHEM equations, the qubit coupling is made with $S_{(m=1)}=\sigma_z$, representing non-Markovian dephasing with strength $\lambda_{(m=1)}=\lambda=0.05$ at temperature $\beta^{-1}=20$. We choose the 
remaining parameters to be similar to those in Ref.(\cite{Vijay:12}):
 a total relaxation rate $r_d=0.05$ that includes the reservoir induced damping $\kappa\mathcal{D}[X]$, and measurement parameters $|\alpha|\chi/\sqrt{\kappa}=0.36$, $\eta=1$, and $\phi-\arg(\alpha)=-\pi/2$. The qubit is driven at Rabi frequency $\Omega_R=3$ and the power spectrum is obtained by averaging over $10^5$ trajectories. The spectrum $S(\omega)$ plotted in 
Figure~\ref{Pspectrum} clearly shows
a shift of 
the peak from $\Omega_R$ and a broadening, both of which are monotonically enhanced as we raise non-Markovian effects by lowering $\gamma_{(m=1)}=\gamma$ from $50$ to $5$. We also observed a shift by keeping $\gamma$ constant and increasing $\lambda$.
In general, the shift can be blue or red depending on parameter values. 

More detailed discussion and analysis of the dependence of the detector spectrum on temperature and on the spectral density parameters $\lambda$ and $\gamma$ are given in appendix D.

The spectrum peak shift is due to the environmentally induced lamb shift \cite{Ramsay}.
In principle the power spectrum can be obtained by 
evaluating the time autocorrelation function from a single experimental run \cite{PSD}, while a time discrete tomographic approach such as in Ref.\cite{Ramsay} asks for many runs. The nonzero offset of the spectrum in Figure (\ref{Pspectrum}) is due to the unavoidable detector noise \cite{Mollow}. 
Such a weak spectroscopy experiment may be carried out in a coupled quantum dot-resonator system, where it would allow probing of 
non-Markovian contributions to decoherence \cite{Frey:12}.

\textit{Discussion and Outlook --}  We presented a microscopic description of CQED-based continuous measurement of a quantum system coupled to a bosonic environment 
that resulted in a stochastic hierarchy of equations of motion (SHEM). While the explicit analysis was presented 
here for an environment of local baths with Drude-Lorentz spectral density, it is straightforward to relax  these assumptions and derive the SHEM for very general spectral densities\cite{Strumpfer,Jin:07, Ma:12}.
Spin environments present more significant challenges for this analysis \cite{Spin} and constitute another interesting direction for future research.

The theory presented here is for a single quantum system. Working with an ensemble allows a higher signal to noise ratio \cite{Silberfarb:05}.  However the associated inhomogeneous broadening as well as coupling between individual ensemble members, 
induced virtually by the dispersive transformation Eq.~(\ref{Hsc}), introduces additional dephasing effects. 

The fundamental nature of the observable engineering developed in this work has revealed a novel spectroscopic capability of quantum weak measurement to probe the non-Markovian nature of the environment in an open quantum system.
This suggests that a new generation of spectroscopy 
techniques might be possible with this approach.   Measurement-based feedback control \cite{Wiseman:book} is another area where SHEM can lead to significant developments.  Feedback control of quantum systems has thus far been limited to systems with weak and Markovian decoherence dynamics. 
The SHEM approach provides the necessary tools to develop feedback control now for open quantum systems with non-Markovian and non-perturbative decoherence effects.  It thereby opens a route to use of quantum feedback control for mitigating effects of non-Markovian quantum noise and a performance comparison with methods such as dynamical decoupling \cite{Lidar:book}.  

We acknowledge funding from DARPA under the QUEST program.  This research was also supported in part by the National Science Foundation under Grant No. NSF PHY11-25915.  AS and KBW thank the Kavli Institute for Theoretical Physics for hospitality.

\onecolumngrid
\section{Appendix}
\subsection{A. System and Cavity Combined decoherence dynamics}

We consider the following Hamiltonian for the system and cavity interactions with the surrounding environment $E$ and electromagnetic reservoir $R$, all in the dispersive regime (see Eqs. (2,3,7) in the paper):
\begin{equation}
H_{SCER}^D=H_{SC}^D+H^D_{drive}+\sum_{m\ge 0,r}g_{mr}(b_{mr}+b_{mr}^\dagger ) F_m
\end{equation} 
We employ a Drude-Lorentz spectral density for both $E$ and $R$, i.e., $J_m(\omega)=2\lambda_m\gamma_m\frac{\omega}{\omega^2+\gamma_m^2}$,
where  $\lambda_m$ and $\gamma_m$ represent the coupling strength and cut-off frequency for the environment mode $m$.
The correlation function corresponding to the Drude-Lorentz model is 
\begin{equation}
\langle \mathcal{T} \tilde{b}_m(t)\tilde{b}_m(0)\rangle_{E\text{ or }R}=\sum_{a=0}^\infty c_{ma}\exp(-\gamma_{ma}t)\label{corr-func}
\end{equation}
where $\gamma_{m0}=\gamma_{m}$ and for $a\ge 1$, $\gamma_{ma}=2\pi a/\beta$ are Matsubara frequencies.
The coefficients $c_{ma}$ are
\begin{equation}
c_{ma}=\Bigg {\{} \begin{tabular}{l}
$\frac{\lambda_m\gamma_m}{2}[cot(\frac{\beta\gamma_m}{2})-i]$,    $a=0$\\ \\
$(\frac{2\lambda_m}{\beta})\frac{\gamma_m\gamma_{ma}}{(\gamma_{ma}^2-\gamma_{m}^2)}$,    $a\ge 1$
\end{tabular}
\end{equation}
The summation Eq.~(\ref{corr-func}) will be truncated at a number $L$ large enough that $\gamma_{mL}\exp( -\gamma_{mL} t)\approx \delta(t)$.

The combined system and cavity dynamics has an exact solution given by the HEOM. The HEOM is a special case where the total bath consists of a non-Markovian environment and a Markovian reservoir. In this situation we may use the hybrid Markov-HEOM equations developed in Ref. \cite{Kreisbeck}.  These yield the following equations for the system-cavity dynamics, which are valid for any temperature:
\begin{eqnarray}
\frac{d\sigma_{{\bf{n}}(t)}}{dt}=(\mathcal{L}_{SC}+\mathcal{L}_{leak}-\nu_{\bf{n}})\sigma_{{\bf{n}}}(t)-\sum_{m\ge 1}\Gamma_m[F_m,[F_m,\sigma_{{\bf{n}}}(t)]]\hspace{1.7 in}\notag\\
-i\sum_{m\ge 1}\sum_{a=0}^{L}[F_m,\sigma_{n_{ma}+1}(t)]-i\sum_{m\ge 1}\sum_{a=0}^{L}n_{ma}(c_{ma}F_m\sigma_{n_{ma}-1}(t)-c_{ma}^*\sigma_{n_{ma}-1}(t)F_m),\label{HEOM-app}
\end{eqnarray}
with $\nu_{\bf{n}}=\sum_{m\ge 1}\sum_{a=0}^{L}n_{ma}\gamma_{ma}$, $\Gamma_m=(1/\beta\gamma_{m0}-i/2)\lambda_m-\sum_{a=0}^Lc_{ma}/\gamma_{ma}$ and $\mathcal{L}_{SC}.=-i[H_{SC}^D+H^D_{drive},.]+\kappa\mathcal{D}[X].$. The second term $\kappa\mathcal{D}[X]$ is the Purcell type of system decoherence modification and is also a part of the measurement back-action. The superoperator $\mathcal{L}_{leak}=\kappa\mathcal{D}[a(1+\Lambda)]$ denotes the modified cavity leakage process. The subscript $\bf{n}$ is a matrix of indices $n_{ma}\ge 0$ and the index $n_{ma}\pm 1$ denotes increase or decrease of the index $n_{ma}$.  The system-cavity density matrix
corresponds to index zero, $\rho_{SC}=\sigma_{\bf{n=0}}$, the remaining operators $\sigma_{\bf{n}\neq 0}$ constitute a set of  auxiliary Hermitian operators.
These hierarchical equations continue to infinite $n$. In practice, however,
one can truncate them at a finite tier $n^t$, which is given by the condition
\begin{eqnarray}
\Psi=\sum_{m\ge 1}\sum_{a=0}^{L}n^t_{ma}\gamma_{ma}\gg, \omega_{SC}
\end{eqnarray}
where $\omega_{SC}^{-1}$ is the largest time scale in the system-cavity dynamics \cite{Ishizaki:10}. This inequality follows from requiring the approximation $\Psi e^{-\Psi (t-s)}\approx \delta(t-s)$ be satisfied.

\subsection{B. Unravelling the HEOM}

We describe a consistent unravelling of the set of dynamical equations (\ref{HEOM-app}) as a result of continuous homodyne measurement of the cavity mode. Although the HEOM describes a non-Markovian evolution, mathematically it has an inherent Markovian property. Eqs~(\ref{HEOM-app}) are clearly a set of linear equations for an extended variable $\Xi=[\sigma_{{\bf{0}}},\sigma_{{\bf{1}}},...,\sigma_{{\bf{N}}}]$. Therefore, we can express the HEOM in a compact form as $\dot{\Xi}=(\Phi+I_{{\bf{N}}}\otimes\mathcal{L}_{leak}) \Xi$, with a super-operator $\Phi$ representing all other terms in Eq.(\ref{HEOM-app}). The HEOM can then be formally solved as 
\begin{equation}
\Xi(t)=e^{(\Phi+I_{{\bf{N}}}\otimes\mathcal{L}_{leak})t}\Xi(0)=\lim_{n\to \infty}(e^{\Phi t/n}I_{{\bf{N}}}\otimes e^{\mathcal{L}_{leak} t/n})^n\Xi(0)
\label{LTK}
\end{equation}
where the second equality is an application of Lie-Trotter-Kato product formula \cite{Trotter}. Eq.(\ref{LTK}) facilitates a separate treatment of the photon leakage process and the decoherence dynamics. The infinitesimal dynamical map $e^{\Phi t/n}$ is the unobserved part of the system-cavity dynamics while the infinitesimal mapping $e^{\mathcal{L}_{leak} t/n}$ is the one being unravelled by the photo-detector. Homodyne detection is based on mixing the cavity signal with a strong local oscillator with amplitude $\beta$ and phase $\phi$.  We can follow the steps in Ref.\cite{Wiseman:book} and represent the the infinitesimal map $e^{\mathcal{L}_{leak} t/n}$ by a Kraus map
\begin{equation}
e^{\mathcal{L}_{leak} t/n}\approx I+\mathcal{L}_{leak} t/n{\bf{.}}=E_0(t/n){\bf{.}}E_0^\dagger(t/n)+E_1(t/n){\bf{.}}E_1^\dagger(t/n),
\end{equation}
with Kraus operators 
\begin{eqnarray}
E_0(t/n)=\sqrt{t/n}[a(1+\Lambda)+\beta e^{i\phi}]\hspace{3.25 in}\\
E_1(t/n)=1-(t/n)[\beta(ae^{-i\phi}-a^\dagger e^{i\phi})(1+\Lambda)+\frac{1}{2}(a^\dagger(1+\Lambda)+\beta e^{-i\phi})(a(1+\Lambda)+\beta e^{i\phi})].
\end{eqnarray}
The unnormalized superoperator $\mathcal{F}_i{\bf{.}}=E_i(t/n){\bf{.}}E_i^\dagger(t/n)$ refers to no-photon (single photon) detection process for $i=0$ ($i=1$). For a history of detector records $\{i_1,i_2,...\}$, the HEOM Eq.(\ref{LTK}) is unraveled as
\begin{equation}
\Xi_{\{i_1,i_2,...\}}(t)=\lim_{n\to \infty}...e^{\Phi t/n}(I_{N}\otimes \mathcal{F}_{i_2})e^{\Phi t/n}(I_{N}\otimes \mathcal{F}_{i_1})\Xi(0)
\label{LTK2}
\end{equation}
where $\Xi_{\{i_1,i_2,...\}}(t)$ is the unnormalized conditional state variable.

We obtain the continuous limit of the discrete picture in EQ.~(\ref{LTK2}) by using the well-known stochastic description of homodyne detection induced dynamics \cite{Wiseman:book,Gambetta:08}.
For a bare cavity, continuous homodyne detection of the leakage process $\dot{\rho_C}=\kappa\mathcal{D}[a]\rho_C$ is described by the following stochastic differential equations (SDEs) of the photo-detector current 
\begin{eqnarray}
d\mathcal{Q}=2\eta\kappa\langle e^{-i\phi}a+e^{i\phi}a^\dagger\rangle dt+\sqrt{2\eta\kappa}dW \label{current},
\end{eqnarray}
together with the associated conditional state of the cavity
\begin{eqnarray}
d\bar{\rho}_{C}=-i[\omega_c,\bar{\rho}_C]dt+\kappa\mathcal{D}[a]\bar{\rho}_Cdt+\sqrt{2\eta\kappa}\mathcal{H}[e^{-i\phi}a]\bar{\rho}_{C}dW.
\label{bare}
\end{eqnarray}
The solution of the SDE (\ref{bare}) for one particular realization of the Wiener process $dW$ is equivalent to a sequence of super-operations $...\mathcal{F}_{i_3}\mathcal{F}_{i_2}\mathcal{F}_{i_1}$.
Thereby, we arrive at the following stochastic HEOM to describe the homodyne measurement of the system and cavity such that the trajectory  described by Eq.~(\ref{LTK2}) is a solution for one particular photo-detector record:
\begin{eqnarray}
d\sigma_{{\bf{n}}}=(\mathcal{L}_{SC}+\mathcal{L}_{leak}-\nu_{\bf{n}})\sigma_{{\bf{n}}}dt-\sum_{m\ge 1}\Gamma_m[F_m,[F_m,\sigma_{{\bf{n}}}]]dt-i\sum_{m\ge 1}\sum_{a=0}^{L}[F_m,\sigma_{n_{ma}+1}]dt\notag\\
-i\sum_{m\ge 1}\sum_{a=0}^{L}n_{ma}(c_{ma}F_m\sigma_{n_{ma}-1}-c_{ma}^*\sigma_{n_{ma}-1}F_m)dt+\sqrt{2\eta\kappa}\mathcal{H}[e^{-i\phi}a(1+\Lambda)]\sigma_{{\bf{n}}}dW.\label{HEOMapp}
\end{eqnarray}

The detector current can be written accordingly as

\begin{eqnarray}
d\mathcal{Q}=2\eta\kappa\langle (1+\Lambda)(e^{-i\phi}a+e^{i\phi}a^\dagger)\rangle dt+\beta\sqrt{2\eta\kappa}dW. \label{current2}
\end{eqnarray}

\subsection{C. Cavity Mode Elimination and the general SHEM}

The next step is to eliminate the cavity mode in the parameter regime where the cavity state reaches to equilibrium with the system state in a negligible time.
In another words, the system has an adiabatic evolution in compare to the cavity dynamics. Such behavior can be obtained with a relatively high leakage (low finesse) cavity.
In the following 
we use the standard approach as described in Refs.~\cite{Doherty:99,Hutchison:09} to eliminate the cavity mode. 

Under cavity driving by the field $\mathcal{E}_pe^{-i\omega_p t}$, it is biased to the coherent steady-state $|\alpha\rangle=|-i\mathcal{E}_p/(i\Delta+\kappa)\rangle$, where $\Delta=\omega_c-\omega_p$. Elimination of the cavity mode then proceeds as follows.\\

1- Write Eqs. (\ref{HEOMapp},\ref{current2}) in the frame rotating with the drive frequency $\omega_p$.

2- Project the cavity to the ground state by the transformation $\rho_c\rightarrow D(-\alpha)\rho_cD(\alpha)$ where $D(\alpha)$ is the displacement operator $D(\alpha)=\exp(\alpha a^\dagger-\alpha^* a)$.

3- Represent the system-cavity density matrix as $\sigma_{\bf{n}}=\sum_{lk}\sigma^{\bf{n}}_{lk}|l\rangle\langle k|$, where $|l\rangle$ is the cavity $l$ photon state in the displaced framework, $D(-\alpha){\bf{.}}D(\alpha)$, and $\sigma^{\bf{n}}_{lk}$ is the corresponding system operator. Expand the density matrix $\rho_{SC}$ to the second order of the perturbative parameters $\epsilon=\frac{1}{\kappa}\max  \{(|O_S|+\Delta)(1+|\alpha|^2),\sum_{m\ge 1}\Gamma_m||Q_m||,\sum_{m\ge 1}|\alpha|^2\lambda||Q_m||\}$. The high leakage condition corresponds then to $\epsilon\ll 1$.

We begin by writing the system-cavity state $\rho_{SC}$ and the auxiliary states $\sigma_{\bf{n}}$ as 
\begin{eqnarray}
\sigma_{\bf{n}}=\sigma^{\bf{n}}_{00}|0\rangle\langle 0|+(\sigma^{\bf{n}}_{10}|1\rangle\langle 0|+h.c.)+\sigma^{\bf{n}}_{11}|1\rangle\langle 1|+(\sigma^{\bf{n}}_{20}|2\rangle\langle 0|+h.c.) \hspace{.5 in} \label{psol2}
\end{eqnarray}

We assume the matrix elements $\sigma^{\bf{n}}_{lk}$ scale as $\epsilon^{l+k}$, consistent with the following perturbative analysis.
We insert the perturbative solution Eq.~(\ref{psol2}) into Eq.~(\ref{HEOMapp}) to find the equations for $d\sigma^{\bf{n}}_{mlk}$. Our ultimate goal is to find the dynamics of the system density matrix $\sigma^{\bf{n}}_{lk}=\sigma^{\bf{n}}_{00}+\sigma^{\bf{n}}_{11}$ that are obtained by solving $\sigma^{\bf{n}}_{20}$ and $\sigma^{\bf{n}}_{10}$ as a function of $\sigma^{\bf{n}}_{00}$ and $\sigma^{\bf{n}}_{11}$. To this end, we shall dissect different terms of the SHEM, Eq.~(\ref{HEOMapp}).  However, showing this detailed analysis we first present the final result for the SHEM at an arbitrary temperature.  In the body of the paper we showed only the high temperature limit of these equations, for both a resonant cavity and resonant drive, $\Delta=0$.
\\

\textit{Stochastic Hierarchy Equations of Motion:}

\begin{eqnarray}
d\sigma^{{\bf{n}}}&&=\mathcal{L}_{S}^{\bf{n}}[\sigma^{{\bf{n}}}]dt-\nu_{\bf{n}}\sigma^{{\bf{n}}}dt+\frac{(\kappa+\nu_{\bf{n}})|\alpha|^2}{(\kappa+\nu_{\bf{n}})^2+\Delta^2}\mathcal{D}[O_S]\sigma^{{\bf{n}}} dt+\frac{i\Delta|\alpha|^2}{(\kappa+\nu_{\bf{n}})^2+\Delta^2}[O^2_S,\sigma^{{\bf{n}}}] dt\notag\\
&&+\sum_{m\ge 1}\Gamma_m\mathcal{D}[\tilde{F}_m]\sigma^{{\bf{n}}}dt-i\sum_{m\ge 1}\sum_{a=0}^{L}[\tilde{F}_m,\sigma_{n_{ma}+1}]dt-i\sum_{m\ge 1}\sum_{a=0}^{L}n_{ma}(c_{ma}(\tilde{F}_m)\sigma_{n_{ma}-1}-c_{ma}^*\sigma_{n_{ma}-1}(\tilde{F}_m))dt\notag\\
&&-\sqrt{2\eta\kappa}\mathcal{H}[e^{-i\phi}\frac{\alpha}{\kappa+i\Delta}(i(1+\Lambda)O_S+\kappa\Lambda^2)]\sigma^{{\bf{n}}} dW
\end{eqnarray}

\textit{Detector Current:}
\begin{eqnarray}
d\mathcal{Q}=\beta[2\eta|\alpha|\langle \frac{\kappa}{\kappa^2+\Delta^2}(1+\Lambda)[\sin(\theta)O_S+\kappa\cos(\theta)\Lambda\rangle dt+\sqrt{2\eta\kappa}dW]
\end{eqnarray}
where $\tilde{F}_m=\tilde{S}_m+Q_m|\alpha|^2$ and $\theta=\phi-\arg(\alpha)-\arctan(\Delta/\kappa)$.

As an extension of this analysis, it would be interesting to consider a cavity with a higher quality factor $Q=\omega_c/\kappa$ and to use the the cavity mode elimination technique developed in Ref. \cite{Gambetta:08}. In the latter case, we can expect reach a higher signal to noise ratio.

\subsection{D. Adiabatic Elimination Calculations}

\subsection{Part I}
First we calculate the action of $(\mathcal{L}_{SC}+\mathcal{L}_{leak})$ on $\sigma=\sigma_{00}|0\rangle\langle 0|+\sigma_{10}|1\rangle\langle 0|+\sigma_{10}^\dagger|0\rangle\langle 1|+\sigma_{11}|1\rangle\langle 1|$.

\begin{eqnarray}
[\mathcal{L}_{SC}+\mathcal{L}_{leak}]\sigma=\mathcal{L}_{S}\sigma-i[(\mathcal{E}_{p}a^\dagger+\mathcal{E}_{p}^*a)(1+\Lambda),\sigma]-i[a^\dagger a(O_{S}+\Delta),\sigma]+\kappa\mathcal{D}[a(1+\Lambda)]\sigma
\end{eqnarray}
where $\Lambda=\frac{1}{2}[X^\dagger,X]$ and
\begin{eqnarray}
\mathcal{L}_{S}\sigma=-i[H_S^D- \sum_q\mathcal{E}_qe^{-i\omega_{q} t}X^\dagger- \sum_q\mathcal{E}^*_qe^{i\omega_{q} t}X,\sigma]+\kappa\mathcal{D}[X]\sigma
\end{eqnarray}

After the transformation $\sigma\rightarrow D(-\alpha)\sigma D(\alpha)$   
\begin{eqnarray}
[\mathcal{L}_{SC}+\mathcal{L}_{leak}]\sigma\rightarrow \hspace{4.3 in} \notag\\
\mathcal{L}_{S}^{\bf{n}}\sigma-i[(\mathcal{E}_{p}a^\dagger+\mathcal{E}_{p}^*a)(1+\Lambda),\sigma]-i[(a^\dagger a+\alpha^* a+\alpha a^\dagger)(O_S+\Delta),\sigma]+\kappa\mathcal{D}[(a+\alpha)(1+\Lambda)]\sigma
\end{eqnarray}
for $\mathcal{L}_{S}^{\bf{n}}\sigma=\mathcal{L}_{S}\sigma-i(\mathcal{E}_{p}\alpha^*+\mathcal{E}_{p}^*\alpha)[\Lambda,\sigma]-i|\alpha|^2[(O_S+\Delta),\sigma]$. Next we apply the superoperator $\mathcal{L}_{SC}+\mathcal{L}_{leak}$ on all terms of $\sigma$:
\vspace{.3 in}

A. Term $\sigma_{00}|0\rangle\langle 0|$:
\begin{eqnarray}
\mathcal{L}_{S}^{\bf{n}}[\sigma_{00}]|0\rangle\langle 0|-((i\mathcal{E}_{p}(1+\Lambda)+i\alpha (O_S+\Delta)+\kappa\alpha(1+2\Lambda))\sigma_{00}|1\rangle\langle 0|+h.c.)
\end{eqnarray}

B. Term $\sigma_{10}|1\rangle\langle 0|$:
 \begin{eqnarray}
 \mathcal{L}_{S}^{\bf{n}}[\sigma_{10}]|1\rangle\langle 0|-(i\mathcal{E}^*_{p}(1+\Lambda)+i\alpha^* (O_S+\Delta))\sigma_{10}|0\rangle\langle 0|+\sigma_{10}(i\mathcal{E}^*_{p}(1+\Lambda)+i\alpha^* (O_S+\Delta))|1\rangle\langle 1|\notag\\
 +\kappa\alpha^* \sigma_{10} (1+2\Lambda)|0\rangle\langle 0|-i (O_S+\Delta)\sigma_{10}|1\rangle\langle 0|-\kappa(1+2\Lambda)\sigma_{10}|1\rangle\langle 0|-\kappa\alpha^* \sigma_{10}(1+2\Lambda)|1\rangle\langle 1|\notag\\
\sqrt{2}\sigma_{10} (i\mathcal{E}^*(1+\Lambda)+i\alpha (O_S+\Delta)-\kappa\alpha^*(1+\Lambda)^2)|0\rangle\langle 2|
\end{eqnarray}

C. Term $\sigma_{11}|1\rangle\langle 1|$:
 \begin{eqnarray}
 \mathcal{L}_{S}^{\bf{n}}[\sigma_{11}]|1\rangle\langle 1|+(\sigma_{11}(i\mathcal{E}_{p}(1+\Lambda)+i\alpha (O_S+\Delta))|1\rangle\langle 0|+\kappa\alpha(1+2\Lambda)\sigma_{11}|1\rangle\langle 0|+h.c.)\notag\\
 -i[(O_S+\Delta),\sigma_{11}]|1\rangle\langle 1|+2\kappa(1+\Lambda)\sigma_{11}(1+\Lambda)|0\rangle\langle 0|-\kappa\{\sigma_{11},1+2\Lambda\}|1\rangle\langle 1|
\end{eqnarray}   

D. Term $\sigma_{20}|2\rangle\langle 0|$:
 \begin{eqnarray}
 \mathcal{L}_{S}^{\bf{n}}[\sigma_{20}]|2\rangle\langle 0|-i\sqrt{2}(\mathcal{E}_{p}^*(1+\Lambda)+\alpha^* (O_S+\Delta))\sigma_{20}|1\rangle\langle 0|-i2\alpha^* (O_S+\Delta)\sigma_{20}|2\rangle\langle 0|\notag\\+2\sqrt{2}\kappa\alpha^*(1+\Lambda)\sigma_{20}(1+\Lambda)|1\rangle\langle 0|-\sqrt{2}\kappa\alpha^*(1+\Lambda)^2\sigma_{20}|1\rangle\langle 0|-2\kappa(1+\Lambda)^2\sigma_{20}|2\rangle\langle 0|\notag\\
\end{eqnarray}   

\subsection{Part II}
     
The next term is $D(-\alpha)\mathcal{H}[e^{-i\phi}a(1+\Lambda)]\sigma D(\alpha)=\mathcal{H}[e^{-i\phi}(a+\alpha)(1+\Lambda)]\sigma$  
\vspace{.3 in}

A. Term $\sigma_{00}|0\rangle\langle 0|$:
\begin{eqnarray}
(e^{-i\phi}\alpha\Lambda\sigma_{00}+e^{i\phi}\alpha^*\sigma_{00}\Lambda)|0\rangle\langle 0|\hspace{4in}\notag \\
-Tr[e^{-i\phi}\alpha\Lambda(\sigma_{00}+\sigma_{11})+e^{i\phi}\alpha^*(\sigma_{00}+\sigma_{11})\Lambda+e^{-i\phi}(1+\Lambda)\sigma_{10}+e^{i\phi}\sigma_{10}^\dagger(1+\Lambda)]\sigma_{00}|0\rangle\langle 0|
\end{eqnarray}   

B. Term $\sigma_{10}|1\rangle\langle 0|$:
\begin{eqnarray}
e^{-i\phi}(1+\Lambda)\sigma_{10}|0\rangle\langle 0|+e^{-i\phi}\Lambda\sigma_{10}\alpha |1\rangle\langle 0|+e^{i\phi}\sigma_{10}\Lambda\alpha^* |1\rangle\langle 0|\hspace{2.5in}\notag\\
-Tr[e^{-i\phi}\alpha\Lambda(\sigma_{00}+\sigma_{11})+e^{i\phi}\alpha^*(\sigma_{00}+\sigma_{11})\Lambda+e^{-i\phi}(1+\Lambda)\sigma_{10}+e^{i\phi}\sigma_{10}^\dagger(1+\Lambda)]\sigma_{10}|1\rangle\langle 0|
\end{eqnarray}   

C. Term $\sigma_{11}|1\rangle\langle 1|$:
\begin{eqnarray}
(e^{-i\phi}\alpha\Lambda\sigma_{11}+e^{i\phi}\alpha^*\sigma_{11}\Lambda)|1\rangle\langle 1|+(e^{-i\phi}(1+\Lambda)\sigma_{11}|0\rangle\langle 1|+e^{i\phi}\sigma_{11}(1+\Lambda)|1\rangle\langle 0|)\hspace{1in}\notag \\
-Tr[e^{-i\phi}\alpha\Lambda(\sigma_{00}+\sigma_{11})+e^{i\phi}\alpha^*(\sigma_{00}+\sigma_{11})\Lambda+e^{-i\phi}(1+\Lambda)\sigma_{10}+e^{i\phi}\sigma_{10}^\dagger(1+\Lambda)]\sigma_{11}|1\rangle\langle 1|
\end{eqnarray}   

D. Term $\sigma_{20}|2\rangle\langle 0|$:
\begin{eqnarray}
\sqrt{2}e^{-i\phi}(1+\Lambda)\sigma_{20}|1\rangle\langle 0|+(e^{-i\phi}\alpha (1+\Lambda)\sigma_{20}+e^{i\phi}\alpha^*\sigma_{20}(1+\Lambda))|2\rangle\langle 0|\hspace{1in}\notag \\
-Tr[e^{-i\phi}\alpha\Lambda(\sigma_{00}+\sigma_{11})+e^{i\phi}\alpha^*(\sigma_{00}+\sigma_{11})\Lambda+e^{-i\phi}(1+\Lambda)\sigma_{10}+e^{i\phi}\sigma_{10}^\dagger(1+\Lambda)]\sigma_{11}|2\rangle\langle 0|
\end{eqnarray}

\subsection{Part III}

The next term requiring the perturbative treatment is $[F_m,\sigma]$ for $F_m=\tilde{S}_m+Q_ma^\dagger a$ with operators $\tilde{S}_m=S_m-\frac{1}{2}\{X^\dagger X,S_m\}+X^\dagger S_mX$ and $Q_m=\mathcal{D}[X]S_m+\mathcal{D}[X^\dagger] S_m$.

\begin{eqnarray}
[\tilde{S}_m+Q_m|\alpha|^2,\sigma]+[Q_m(a^\dagger a+a\alpha^*+a^\dagger \alpha),\sigma]
\end{eqnarray}

A. Term $\sigma_{00}|0\rangle\langle 0|$:
\begin{eqnarray}
[\tilde{S}_m+Q_m|\alpha|^2,\sigma_{00}]|0\rangle\langle 0|+Q_m \alpha\sigma_{00}|1\rangle\langle 0|-\sigma_{00}Q_m \alpha^*|0\rangle\langle 1|
\end{eqnarray}   

B. Term $\sigma_{10}|1\rangle\langle 0|$:
\begin{eqnarray}
[\tilde{S}_m+Q_m|\alpha|^2,\sigma_{10}]|1\rangle\langle 0|+Q_m \sigma_{10}|1\rangle\langle 0|+\alpha^*Q_m\sigma_{10}|0\rangle\langle 0|-\sigma_{10}Q_m \alpha^*|1\rangle\langle 1|+\sqrt{2}\alpha Q_m\sigma_{10}|2\rangle\langle 0|
\end{eqnarray}   

C. Term $\sigma_{11}|1\rangle\langle 1|$:
\begin{eqnarray}
[\tilde{S}_m+Q_m|\alpha|^2,\sigma_{11}]|1\rangle\langle 1|+[Q_m,\sigma_{11}]|1\rangle\langle 1|+\alpha^* Q_m \sigma_{11}|0\rangle\langle 1|-\sigma_{11}Q_m \alpha |1\rangle\langle 0|
\end{eqnarray}   

D. Term $\sigma_{20}|2\rangle\langle 0|$:
\begin{eqnarray}
[\tilde{S}_m+Q_m|\alpha|^2,\sigma_{20}]|2\rangle\langle 0|+2Q_m\sigma_{20}|2\rangle\langle 0|+\alpha^*\sqrt{2} Q_m \sigma_{20}|1\rangle\langle 0|
\end{eqnarray}

\subsection{Part IV}

We also have the double commutator operation $[F_m,[F_m,\sigma]]$ that is expanded to the second order of $Q_m$

\begin{eqnarray}
[\tilde{S}_m,\sigma]+[Q_m|\alpha|^2,\sigma]+[Q_m(a^\dagger a+a\alpha^*+a^\dagger \alpha),\sigma]
\end{eqnarray}

A. Term $\sigma_{00}|0\rangle\langle 0|$:
\begin{eqnarray}
\big ([\tilde{S}_m,[\tilde{S}_m,\sigma_{00}]]+[\tilde{S}_m,[Q_m|\alpha|^2,\sigma_{00}]]+[Q_m|\alpha|^2,[\tilde{S}_m,\sigma_{00}]]\big )|0\rangle\langle 0|+[Q_m \alpha [\tilde{S}_m,\sigma_{00}]|1\rangle\langle 0|-[\tilde{S}_m,\sigma_{00}]Q_m \alpha^*|0\rangle\langle 1|\notag\\
+[\tilde{S}_m,Q_m \alpha\sigma_{00}]|1\rangle\langle 0|-[\tilde{S}_m,\sigma_{00}Q_m \alpha^*]|0\rangle\langle 1|
\end{eqnarray}   

B. Term $\sigma_{10}|1\rangle\langle 0|$:
\begin{eqnarray}
[\tilde{S}_m+Q_m|\alpha|^2,[\tilde{S}_m,\sigma_{10}]]|1\rangle\langle 0|+Q_m [\tilde{S}_m,\sigma_{10}]|1\rangle\langle 0|+\alpha^*Q_m[\tilde{S}_m,\sigma_{10}]|0\rangle\langle 0|-[\tilde{S}_m,\sigma_{10}]Q_m \alpha^*|1\rangle\langle 1|+\sqrt{2}\alpha Q_m[\tilde{S}_m,\sigma_{10}]|2\rangle\langle 0|\notag\\
+[\tilde{S}_m,[Q_m|\alpha|^2,\sigma_{10}]]|1\rangle\langle 0|+[\tilde{S}_m,Q_m \sigma_{10}]|1\rangle\langle 0|+[\tilde{S}_m,\alpha^*Q_m\sigma_{10}]|0\rangle\langle 0|-[\tilde{S}_m,\sigma_{10}Q_m \alpha^*]|1\rangle\langle 1|+\sqrt{2}[\tilde{S}_m,\alpha Q_m\sigma_{10}]|2\rangle\langle 0|
\end{eqnarray}   

C. Term $\sigma_{11}|1\rangle\langle 1|$:
\begin{eqnarray}
[\tilde{S}_m+Q_m|\alpha|^2,[\tilde{S}_m,\sigma_{11}]]|1\rangle\langle 1|+[Q_m,[\tilde{S}_m,\sigma_{11}]]|1\rangle\langle 1|+\alpha^* Q_m [\tilde{S}_m,\sigma_{11}]|0\rangle\langle 1|-[\tilde{S}_m,\sigma_{11}]Q_m \alpha |1\rangle\langle 0|\\
+[\tilde{S}_m,[Q_m|\alpha|^2,\sigma_{11}]]|1\rangle\langle 1|+[\tilde{S}_m,[Q_m,\sigma_{11}]]|1\rangle\langle 1|+[\tilde{S}_m,\alpha^* Q_m \sigma_{11}]|0\rangle\langle 1|-[\tilde{S}_m,\sigma_{11}Q_m \alpha] |1\rangle\langle 0|
\end{eqnarray}   

D. Term $\sigma_{20}|2\rangle\langle 0|$:
\begin{eqnarray}
[\tilde{S}_m+Q_m|\alpha|^2,[\tilde{S}_m,\sigma_{20}]]|2\rangle\langle 0|+2Q_m[\tilde{S}_m,\sigma_{20}]|2\rangle\langle 0|+\alpha^*\sqrt{2} Q_m [\tilde{S}_m,\sigma_{20}]|1\rangle\langle 0|\notag\\
+[\tilde{S}_m,[Q_m|\alpha|^2,\sigma_{20}]]|2\rangle\langle 0|+2[\tilde{S}_m,Q_m\sigma_{20}]|2\rangle\langle 0|+\alpha^*\sqrt{2} [\tilde{S}_m,Q_m \sigma_{20}]|1\rangle\langle 0|
\end{eqnarray}

\subsection{Part V} 

The next term is $\{F_m,\sigma\}$.
\\

A. Term $\sigma_{00}|0\rangle\langle 0|$:
\begin{eqnarray}
\{\tilde{S}_m+Q_m|\alpha|^2,\sigma_{00}\}|0\rangle\langle 0|+\{Q_m \alpha\sigma_{00}|1\rangle\langle 0|+\sigma_{00}Q_m \alpha^*|0\rangle\langle 1|)\}
\end{eqnarray}   

B. Term $\sigma_{10}|1\rangle\langle 0|$:
\begin{eqnarray}
\{\tilde{S}_m+Q_m|\alpha|^2,\sigma_{10}\}|1\rangle\langle 0|+Q_m \sigma_{10}|1\rangle\langle 0|+\alpha^*Q_m\sigma_{10}|0\rangle\langle 0|+\sigma_{10}Q_m \alpha^*|1\rangle\langle 1|+\sqrt{2}\alpha Q_m\sigma_{10}|2\rangle\langle 0|
\end{eqnarray}   

C. Term $\sigma_{11}|1\rangle\langle 1|$:
\begin{eqnarray}
\{\tilde{S}_m+Q_m|\alpha|^2,\sigma_{11}\}|1\rangle\langle 1|+\{Q_m,\sigma_{11}\}|1\rangle\langle 1|+\alpha^* Q_m \sigma_{11}|0\rangle\langle 1|+\sigma_{11}Q_m \alpha |1\rangle\langle 0|
\end{eqnarray}   

D. Term $\sigma_{20}|2\rangle\langle 0|$:
\begin{eqnarray}
\{\tilde{S}_m+Q_m|\alpha|^2,\sigma_{20}\}|2\rangle\langle 0|+2Q_m\sigma_{20}|2\rangle\langle 0|+\alpha^*\sqrt{2} Q_m \sigma_{20}|1\rangle\langle 0|
\end{eqnarray}

\subsection{Part VI} 
Wrapping up the above calculations in parts I-V, we find the following hierarchical equations for different components of $\rho_{SC}$ and $\sigma_{{\bf{n}}}$ 
 
A. $|0\rangle\langle 0|$ component:
\begin{eqnarray}
d\sigma^{{\bf{n}}}_{00}&&=\mathcal{L}_{S}^{\bf{n}}[\sigma^{{\bf{n}}}_{00}]dt-\nu_{\bf{n}}\sigma^{{\bf{n}}}_{00}dt-i(\mathcal{E}^*_{p}\Lambda+\alpha^*O_S)\sigma^{{\bf{n}}}_{10}dt+i\sigma^{{\bf{n}}\dagger}_{10}(\mathcal{E}_{p}\Lambda+\alpha O_S) dt+2\kappa\alpha^*\sigma^{{\bf{n}}}_{10}\Lambda+2\kappa\alpha\Lambda\sigma^{{\bf{n}}\dagger}_{10}\notag\\
&&+2\kappa(1+\Lambda)\sigma^{{\bf{n}}}_{11}(1+\Lambda)+\sum_{m\ge 1}\Gamma_m\mathcal{D}[\tilde{F}_m]\sigma^{{\bf{n}}}_{00}dt+\sum_{m\ge 1}\Gamma_m(\alpha^*Q_m[\tilde{S}_m,\sigma^{{\bf{n}}}_{10}]-\alpha[\tilde{S}_m,\sigma^{\dagger{\bf{n}}}_{10}]Q_m+h.c.) dt\notag\\
&&-i\sum_{m\ge 1}\sum_{a=0}^{L}\big ([\tilde{F}_m,\sigma_{00}^{{n}_{ma}+1}+n_{ma}Re(c_{ma})\sigma_{00}^{{n}_{ma}-1}]+\alpha^*Q_m(\sigma_{10}^{{n}_{ma}+1}+n_{ma}Re(c_{ma})\sigma_{10}^{{n}_{ma}-1})\notag\\
&&-\alpha (\sigma_{10}^{\dagger{n}_{ma}+1}+n_{ma}Re(c_{ma})\sigma_{10}^{{n}_{ma}-1})Q_m\big )dt \notag\\
&&-\sum_{m\ge 1} \{\tilde{F}_m,n_{ma}\sigma_{00}^{{n}_{ma}-1}\}+\alpha^*Q_mn_{ma}\sigma_{10}^{{n}_{ma}-1}+\alpha n_{ma}\sigma_{10}^{{\dagger{n}_{ma}-1}} Q_m\notag \\
&&+\sqrt{2\eta\kappa}(e^{-i\phi}\alpha\Lambda\sigma^{{\bf{n}}}_{00}+e^{i\phi}\alpha^*\sigma^{{\bf{n}}}_{00}\Lambda+e^{-i\phi}(1+\Lambda)\sigma^{{\bf{n}}}_{10}+e^{i\phi}\sigma^{\dagger{\bf{n}}}_{10}(1+\Lambda))dW\notag\\
&&-\sqrt{2\eta\kappa}Tr[e^{-i\phi}\alpha\Lambda\sigma^{{\bf{n}}}_{00}+e^{i\phi}\alpha^*\sigma^{{\bf{n}}}_{00}\Lambda+e^{-i\phi}(1+\Lambda)\sigma^{{\bf{n}}}_{10}+e^{i\phi}\sigma^{\dagger{\bf{n}}}_{10}(1+\Lambda)+e^{-i\phi}\alpha\Lambda\sigma^{{\bf{n}}}_{11}+e^{i\phi}\alpha^*\sigma^{{\bf{n}}}_{11}\Lambda]\sigma^{{\bf{n}}}_{00}dW
\end{eqnarray}

B. $|1\rangle\langle 0|$ component:
\begin{eqnarray}
d\sigma^{{\bf{n}}}_{10}&&=\mathcal{L}_{S}^{\bf{n}}[\sigma^{{\bf{n}}}_{10}]dt-\nu_{\bf{n}}\sigma^{{\bf{n}}}_{10}dt-(i\alpha O_S+\kappa\alpha\Lambda-i\Delta\Lambda)\sigma^{{\bf{n}}}_{00}dt-i(O_S+\Delta)\sigma^{{\bf{n}}}_{10}dt-\kappa(1+2\Lambda)\sigma^{{\bf{n}}}_{10}dt\notag\\
&&+\sum_{m\ge 1}\Gamma_m\mathcal{D}[\tilde{F}_m]\sigma^{{\bf{n}}}_{10}dt-\Gamma_mQ_m[\tilde{S}_m,\sigma^{{\bf{n}}}_{10}]-\Gamma_m[\tilde{S}_m,Q_m\sigma^{{\bf{n}}}_{10}]-\Gamma_m\alpha Q_m[\tilde{S}_m,\sigma^{{\bf{n}}}_{00}]-\Gamma_m\alpha[\tilde{S}_m,Q_m\sigma^{{\bf{n}}}_{00}]\notag\\
&&+\sum_{m\ge 1}\Gamma_m\alpha [\tilde{S}_m,\sigma^{{\bf{n}}}_{11}]Q_m+\Gamma_m\alpha[\tilde{S}_m,\sigma^{{\bf{n}}}_{11}Q_m]-\Gamma_m\sqrt{2}\alpha^* Q_m[\tilde{S}_m,\sigma^{{\bf{n}}}_{20}]-\Gamma_m\sqrt{2}\alpha^*[\tilde{S}_m,\sigma^{{\bf{n}}}_{20}Q_m]\notag\\
&&-i\sum_{m\ge 1}\sum_{a=0}^{L}\big ([\tilde{F}_m,\sigma_{10}^{{n}_{ma}+1}+n_{ma}Re(c_{ma})\sigma_{10}^{{n}_{ma}-1}]+\alpha Q_m(\sigma_{00}^{{n}_{ma}+1}+n_{ma}Re(c_{ma})\sigma_{00}^{{n}_{ma}-1})\notag\\
&&+Q_m(\sigma_{10}^{{n}_{ma}+1}+n_{ma}Re(c_{ma})\sigma_{10}^{{n}_{ma}-1})-\alpha (\sigma_{11}^{{n}_{ma}+1}+n_{ma}Re(c_{ma})\sigma_{00}^{{n}_{ma}-1})Q_m\notag\\
&&+\alpha^*\sqrt{2} Q_m(\sigma_{20}^{{n}_{ma}+1}+n_{ma}Re(c_{ma})\sigma_{20}^{{n}_{ma}-1})\big )dt \notag\\
&&-\sum_{m\ge 1} n_{ma}\{\tilde{F}_m,\sigma_{10}^{{n}_{ma}-1}\}+n_{ma}Q_m\sigma_{10}^{{n}_{ma}-1}+\alpha n_{ma} Q_m\sigma_{00}^{{{n}_{ma}-1}}+\alpha n_{ma} \sigma_{11}^{{{n}_{ma}-1}}Q_m+\alpha^* n_{ma}\sqrt{2}Q_m  \sigma_{20}^{{{n}_{ma}-1}}\notag\\
&&\sqrt{2\eta\kappa}(e^{-i\phi}\Lambda\sigma^{{\bf{n}}}_{10}\alpha+e^{i\phi}\sigma^{{\bf{n}}}_{10}\Lambda\alpha^*+\sqrt{2}e^{-i\phi}(1+\Lambda)\sigma^{{\bf{n}}}_{20})dW\notag\\
&&-\sqrt{2\eta\kappa}Tr[e^{-i\phi}\alpha\Lambda\sigma^{{\bf{n}}}_{00}+e^{i\phi}\alpha^*\sigma^{{\bf{n}}}_{00}\Lambda+e^{-i\phi}\sigma^{{\bf{n}}}_{10}+e^{i\phi}\sigma^{^\dagger{\bf{n}}}_{10}+e^{-i\phi}\alpha\Lambda\sigma^{{\bf{n}}}_{11}+e^{i\phi}\alpha^*\sigma^{{\bf{n}}}_{11}\Lambda]\rho_{10}dW
\end{eqnarray}

C. $|1\rangle\langle 1|$ component:
\begin{eqnarray}
d\sigma^{{\bf{n}}}_{11}&&=\mathcal{L}_{S}^{\bf{n}}[\sigma^{{\bf{n}}}_{11}]dt-\nu_{\bf{n}}\sigma^{{\bf{n}}}_{11}dt+(\sigma^{{\bf{n}}}_{10}(i\mathcal{E}_{p}^*\Lambda+i\alpha^* O_S)+h.c.)dt-\kappa\{\sigma^{{\bf{n}}}_{11},1+2\Lambda\}\notag\\
&&+\sum_{m\ge 1}\Gamma_m\mathcal{D}[\tilde{F}_m]\sigma^{{\bf{n}}}_{11}dt-\sum_{m\ge 1}\Gamma_m(\alpha^*[\tilde{S}_m,\sigma^{{\bf{n}}}_{11}]Q_m+\alpha^*[\tilde{S}_m,\sigma^{{\bf{n}}}_{10}Q_m]+h.c.) dt\notag\\
&&\sum_{m\ge 1}\Gamma_m ([Q_m,[\tilde{S}_m,\sigma^{{n}}_{11}]]+[\tilde{S}_m,[Q_m,\sigma^{{n}}_{11}]])-i\sum_{m\ge 1}\sum_{a=0}^{L}\big ([\tilde{F}_m,\sigma_{11}^{{n}_{ma}+1}+Re(\sigma_{11}^{{n}_{ma}-1})]+[Q_m,\sigma_{11}^{{n}_{ma}+1}+Re(\sigma_{11}^{{n}_{ma}-1})]\notag\\
&&-((\sigma_{10}^{{n}_{ma}+1}+Re(\sigma_{10}^{{n}_{ma}-1}))Q_m\alpha^*+Q_m(\sigma_{10}^{\dagger {n}_{ma}+1}+Re(\sigma_{10}^{\dagger {n}_{ma}-1}))\alpha]))\notag\\
&&-\sum_{m\ge 1} n_{ma}\{\tilde{F}_m,\sigma_{11}^{{n}_{ma}-1}\}+n_{ma}\{Q_m,\sigma_{11}^{{n}_{ma}-1}\}+n_{ma}\alpha^*\sigma_{10}^{{n}_{ma}-1} Q_m+n_{ma}\alpha Q_m\sigma_{10}^{{\dagger{n}_{ma}-1}}\notag\\
&&+\sqrt{2\eta\kappa}(e^{-i\phi}\Lambda\sigma^{{\bf{n}}}_{11}\alpha+e^{i\phi}\sigma^{{\bf{n}}}_{11}\Lambda\alpha^*)dW\notag\\
&&-\sqrt{2\eta\kappa}Tr[e^{-i\phi}\alpha\Lambda\sigma^{{\bf{n}}}_{00}+e^{i\phi}\alpha^*\sigma^{{\bf{n}}}_{00}\Lambda+e^{-i\phi}\sigma^{{\bf{n}}}_{10}+e^{i\phi}\sigma^{\dagger{\bf{n}}}_{10}+e^{-i\phi}\alpha\Lambda\sigma^{{\bf{n}}}_{11}+e^{i\phi}\alpha^*\sigma^{{\bf{n}}}_{11}\Lambda]\sigma^{{\bf{n}}}_{11}dW
\end{eqnarray} 

D. $|2\rangle\langle 0|$ component:
\begin{eqnarray}
d\sigma^{{\bf{n}}}_{20}&&=\mathcal{L}_{S}^{\bf{n}}[\sigma^{{\bf{n}}}_{20}]dt-\nu_{\bf{n}}\sigma^{{\bf{n}}}_{20}dt-i2\alpha^*(O_S+\Delta)\sigma^{{\bf{n}}}_{20}dt-\sqrt{2}\kappa\alpha(1+2\Lambda)\sigma^{\dagger{\bf{n}}}_{10}\notag\\
&&+\sum_{m\ge 1}\Gamma_m\mathcal{D}[\tilde{F}_m]\sigma^{{\bf{n}}}_{20}dt+2Q_m[\tilde{S}_m,\sigma^{{\bf{n}}}_{20}]+2[\tilde{S}_m,Q_m\sigma^{{\bf{n}}}_{20}]+\sqrt{2}\alpha Q_m[\tilde{S}_m,\sigma^{{\bf{n}}}_{10}]+\sqrt{2}\alpha [\tilde{S}_m,Q_m\sigma^{{\bf{n}}}_{10}])\big ) dt\notag\\
&&-i\sum_{m\ge 1}\sum_{a=0}^{L}\big ([\tilde{F}_m,\sigma_{20}^{{n}_{ma}+1}+Re(\sigma_{20}^{{n}_{ma}-1})]+[Q_m,\sigma_{20}^{{n}_{ma}+1}+Re(\sigma_{20}^{{n}_{ma}-1})]\notag\\
&&2Q_m((\sigma_{10}^{{n}_{ma}+1}+Re(\sigma_{10}^{{n}_{ma}-1}))+\sqrt{2}\alpha Q_m(\sigma_{10}^{\dagger {n}_{ma}+1}+Re(\sigma_{10}^{\dagger {n}_{ma}-1}))\big )\notag\\
&&-\sum_{m\ge 1} n_{ma} \{\tilde{F}_m,\sigma_{11}^{{n}_{ma}-1}\}+n_{ma}\{Q_m,\sigma_{11}^{{n}_{ma}-1}\}+n_{ma}\alpha^*\sigma_{10}^{{n}_{ma}-1} Q_m+n_{ma}\alpha Q_m\sigma_{10}^{{\dagger{n}_{ma}-1}}\notag\\
&&-\sum_{m\ge 1} n_{ma}\{\tilde{F}_m,\sigma_{20}^{{n}_{ma}-1}\}+2n_{ma}Q_m\sigma_{20}^{{n}_{ma}-1}+\sqrt{2}\alpha n_{ma} Q_m\sigma_{10}^{{n}_{ma}-1}\notag\\
&&+\sqrt{2\eta\kappa}(e^{-i\phi}(1+\Lambda)\sigma^{{\bf{n}}}_{20}+e^{i\phi}\sigma^{\dagger{\bf{n}}}_{20}(1+\Lambda))dW\notag\\
&&-\sqrt{2\eta\kappa}Tr[e^{-i\phi}\alpha\Lambda\sigma^{{\bf{n}}}_{00}+e^{i\phi}\alpha^*\sigma^{{\bf{n}}}_{00}\Lambda+e^{-i\phi}\sigma^{{\bf{n}}}_{10}+e^{i\phi}\sigma^{\dagger{\bf{n}}}_{10}+e^{-i\phi}\alpha\Lambda\sigma^{{\bf{n}}}_{11}+e^{i\phi}\alpha^*\sigma^{{\bf{n}}}_{11}\Lambda]\sigma^{{\bf{n}}}_{20}dW
\end{eqnarray}

In the limit of large $\kappa\gg \{|O_S|(1+|\alpha|^2),\sum_{m\ge 1}\Gamma_m||Q_m||,\sum_{m\ge 1}|\alpha|^2\lambda||Q_m||\}$ the off-diagonal terms $\{\rho_{10},\rho_{20},\sigma^{{\bf{n}}}_{10},\sigma^{{\bf{n}}}_{20}\}$ decay faster than diagonal terms. We can then solve for $\sigma^k_{10}$
\begin{eqnarray}
\sigma^k_{10}=-\frac{1}{\kappa+\nu_{\bf{n}}+i\Delta}(i\alpha O_S+\kappa\alpha\Lambda)\sigma^k_{00}+\sigma^k_{11}\frac{1}{\kappa+\nu_{\bf{n}}+i\Delta}(i\alpha O_S+\kappa\alpha\Lambda)+O(\epsilon^2)
\end{eqnarray}

Now we are ready to put everything together and find the final expression for the system only dynamical equation. We sum up the above equations for $d\sigma^n_{00}$ and $d\sigma^n_{11}$ to find the SHEM:
\begin{eqnarray}
d\sigma^{{\bf{n}}}&&=\mathcal{L}_{S}^{\bf{n}}[\sigma^{{\bf{n}}}]dt-\nu_{\bf{n}}\sigma^{{\bf{n}}}dt+\frac{(\kappa+\nu_{\bf{n}})|\alpha|^2}{(\kappa+\nu_{\bf{n}})^2+\Delta^2}\mathcal{D}[O_S]\sigma^{{\bf{n}}} dt+\frac{i\Delta|\alpha|^2}{(\kappa+\nu_{\bf{n}})^2+\Delta^2}[O^2_S,\sigma^{{\bf{n}}}] dt+\sum_{m\ge 1}\Gamma_m\mathcal{D}[\tilde{F}_m]\sigma^{{\bf{n}}}\notag\\
&&-i\sum_{m\ge 1}\sum_{a=0}^{L}[\tilde{F}_m,\sigma_{n_{ma}+1}]dt-i\sum_{m\ge 1}\sum_{a=0}^{L}n_{ma}(c_{ma}(\tilde{F}_m)\sigma_{n_{ma}-1}-c_{ma}^*\sigma_{n_{ma}-1}(\tilde{F}_m))dt\notag\\
&&-\sqrt{2\eta\kappa}\mathcal{H}[e^{-i\phi}\frac{\alpha}{\kappa+i\Delta}(i(1+\Lambda)O_S+\kappa\Lambda^2)]\sigma^{{\bf{n}}} dW
\end{eqnarray}

The associated detector signal is

\begin{eqnarray}
d\mathcal{Q}&=&\beta[-2\eta\kappa\langle (1+\Lambda)(e^{-i\phi}\rho_{10}+e^{i\phi}\rho_{10}^\dagger)\rangle dt+\sqrt{2\eta\kappa}dW]\notag\\
&=&\beta[2\eta|\alpha|\langle \frac{\kappa}{\kappa^2+\Delta^2}(1+\Lambda)[\sin(\theta)O_S+\kappa\cos(\theta)\Lambda\rangle dt+\sqrt{2\eta\kappa}dW]
\end{eqnarray}
where $\theta=\phi-\arg(\alpha)-\arctan(\Delta/\kappa)$.

\subsection{E. Weak spectroscopy analysis of the system-environment coupling}
\label{weak_spectro_appendix}

\begin{figure}
\centering
\includegraphics[width=0.7\textwidth]{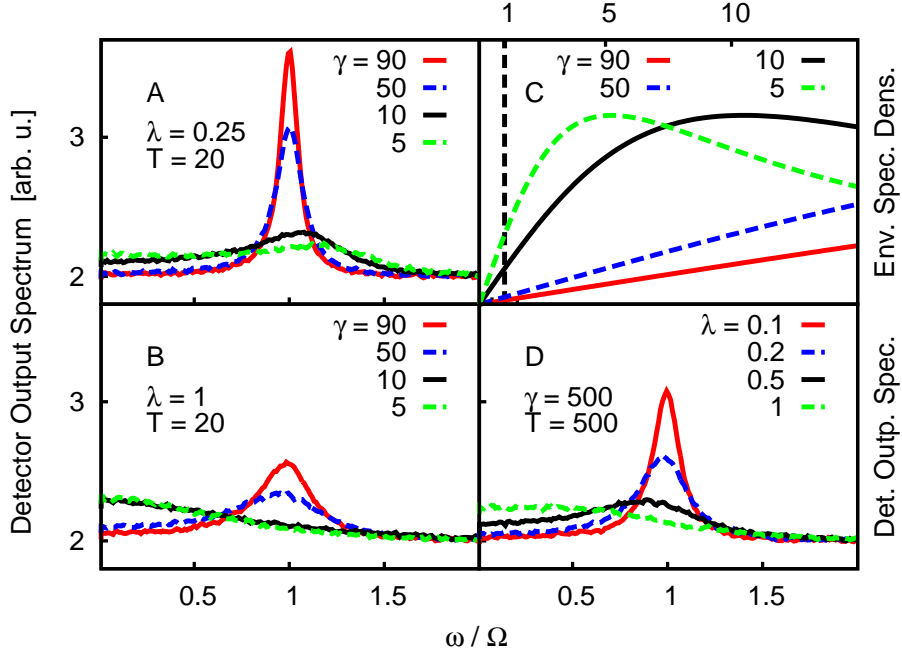}
\caption{A, B, and D: averaged power spectral densities of the detector current for different parameters $\gamma$ and $\lambda$ of the Drude-Lorentz environment spectral density and temperature $T$ (values indicated in the figure).
C: Drude-Lorentz environment spectral densities for four different values of $\gamma$ (values indicated in the figure; all for the same $\lambda$, which is simply a global prefactor). The vertical line at $\omega=1$ marks the transition energy of the two-level system.}
\label{spectra_and_SPD_multiplot}
\end{figure}

The Weak Spectroscopy results presented in the paper (Figure 2) show a clear blue shift of the 
peak that depends on 
the cut-off frequency $\gamma$ of the environment. 
Decreasing the cut-off $\gamma$ leads to a stronger non-Markovianity of the environment.
Thus, the peak shift, which becomes larger for decreasing $\gamma$, could serve as an indicator for the degree of non-Markovianity. 
We show here how the spectra change when we vary not only $\gamma$, but also the parameter $\lambda$ that scales the strength of the coupling to the environment,
as well as the temperature.  
Consider again an environment spectral density of the Drude-Lorentz 
form, $J(\omega)=2\lambda\gamma\frac{\omega}{\omega^2+\gamma^2}$.
This spectral density is shown In 
Figure~\ref{spectra_and_SPD_multiplot} tfor four different values of the cut-off energy $\gamma$.
%
%

Figures~\ref{spectra_and_SPD_multiplot} A and B show the detector spectra for the same values of cut-off $\gamma$ and temperature as previously used for the calculations shown in the paper (Figure 2).
However, the coupling to the environment is made stronger by factor of 5 in Figure~\ref{spectra_and_SPD_multiplot} A, and by a factor of 20 in Figure~\ref{spectra_and_SPD_multiplot} B, i.e., the coupling strength parameter takes the values $\lambda=0.25$ and $\lambda=1$, respectively.  All other parameters of the calculation are the same as for in Figure 2 in the paper.
Figure~\ref{spectra_and_SPD_multiplot} A shows that for a moderate increase in coupling strength, the peaks are broader compared to those in the paper (Figure 2), but the shifts of the peak maxima 
remain roughly the same. We can relate this behavior to the spectral densities for the corresponding $\gamma$ values shown in Figure~\ref{spectra_and_SPD_multiplot} C.
In the neighborhood of the transition energy of the two-level system, marked with the vertical line, the spectral density grows when $\gamma$ is decreased, in a similar fashion as it would grow on increase of $\lambda$ (which is just a prefactor in the Drude-Lorentz spectral density). 
Accordingly, the peaks in the detector spectra broaden when $\gamma$ is decreased or when $\lambda$ is increased, due to the stronger coupling of the two-level transition to the environment. 
Beyond this, however, a shift of the peak in the detector spectrum occurs when the cut-off energy $\gamma$ of the spectral density is reduced and approaches the two-level transition. 
That gives additional information about the character of the coupling to the environment, in particular, of the degree of non-Markovianity.

In Figure~\ref{spectra_and_SPD_multiplot} B, the coupling strength $\lambda$ is further increased and the peaks become even broader. 
Additonally, now for the smaller values of $\gamma$, the spectral densiy in the neighborhood of the two-level transition is so large that the coupling to the environment damps out the oscillations in the detector current, leading to very broad peaks at zero frequency in the detector spectrum for $\gamma = 5, 10$.
This damping of the oscillations 
was already apparent in the rise of the low energy contributions in Figure~\ref{spectra_and_SPD_multiplot} A. 
Further investigations show that increasing the temperature leads to additional broadening of the peaks in the detector spectra.
As an example, in Figure~\ref{spectra_and_SPD_multiplot} D we show detector spectra for a temperature that is higher by a factor of 25.
For this example we choose a large cut-off $\gamma=500$, since for smaller $\gamma$ the peaks would be excessively broad.
Increasing the coupling strength $\lambda$ now shows a similar broadening of the peaks and subsequent damping of the oscillations as seen above for decreasing $\gamma$.
These observations are consistent with the above explanation based on the form of the environment spectral density in relation to the position of the two-level transition.

\subsection{F. Details of the simulations}

The detector output spectra in the paper (Figure 2) and~\ref{spectra_and_SPD_multiplot} were calculated using the general SHEM equations up to the second order in Ito form Eq.(\ref{HEOM-app}).
We calculate the stochastic detector current of the weak measurement over time by solving the SHEM equations with a Runge-Kutta method (of order 1 and 1.5) that gives strong solutions.
The numerical integration part of our simulation uses Fortran 90 routines partly based on routines provided by Ref.\cite{Steck}.
For the calculation of the power spectral density of the detector current we use the Python routine matplotlib.mlab.psd.
To calculate the averaged power spectral densities of the detector current, shown in the paper (Figure 2) and~\ref{spectra_and_SPD_multiplot}, we averaged over $10^5$ individual power spectra of indiviadual stochastic trajectories of the detector current. 
For the two spectra in Figs.~\ref{spectra_and_SPD_multiplot} B for $\gamma=5, 10$, however, where the coupling to the environment is so strong that the oscillations of the detector current are damped out, 24\% (for $\gamma=5$) and 3\% (for $\gamma=10$) of the trajectories diverged and were not taken into account in the averaging.


\begin{thebibliography}{99}
\bibitem{WS} We use the term "Weak Spectroscopy" referring to a weak measurement-based spectroscopy method.
\bibitem{Mollow} We do not observe a Mollow-triplet type sidebands \cite{Roy:11} in the detector spectrum as in the bad cavity limit the homodyne detector only measures the population time correlation $\langle Z(t+\tau)Z(t)\rangle$.
\end{thebibliography}
\end{document}